\documentclass[12pt]{article}
\usepackage{amssymb}
\begin{document}
\def\mkb{\mbox}
\def\beq{\begin{equation}}
\def\eeq{\end{equation}}
\def\beqn{\begin{eqnarray}}
\def\eeqn{\end{eqnarray}}
\def\H{\mathcal{H}}
\def\F{\mathcal{F}}
\def\P{\mathcal{P}}
\def\A{\mathcal{A}}
\def\l{\mathcal{l}}
\def\L{\mathcal{L}}
\def\U{\mathcal{U}}
\def\g{\mathfrak{g}}
\def\C{\mathbb{C}}
\def\R{\mathbb{R}}
\def\K{\mathbb{K}}
\def\N{\mathbb{N}}
\def\Z{\mathbb{Z}}
\thispagestyle{empty}
\begin{center}
{\LARGE{Loday-type Algebras and the}}\\[0.1cm]
{\LARGE{Rota-Baxter Relation}}\\[0.3cm]
\end{center}
\vspace{2cm}
\begin{center}
    K. Ebrahimi-Fard
\footnote{e-mail: fard@th.physik.uni-bonn.de}\\
\vspace{1.5cm}
\vspace{0.2cm}
{\sl Physikalisches Institut der Universit\"at Bonn}\\
{\sl Nu{\ss}allee 12, D--53115 Bonn, Germany}
\end{center}
\vspace{2cm}
\setcounter{page}{1}
\begin{abstract}
In this brief note we would like to report on an observation
concerning the relation between Rota-Baxter operators  and
Loday-type algebras, i.e.~dendriform di- and trialgebras. It is
shown that associative algebras equipped with a Rota-Baxter
operator of arbitrary weight  always give such dendriform
structures.
\end{abstract}
---------------------------------------\\
{\tiny{
\begin{tabular}{ll}
Keywords: & Rota-Baxter relation, dendriform di/trialgebras, classical Yang-Baxter equation,\\
          &associative classical Yang-Baxter equation, minimal subtraction scheme\\
MSC-class:& 16B99;81T15;81T18;17B81\\
\end{tabular}        }}
\newpage
%
\section{Introduction}
Rota-Baxter operators\footnote{We call them here Rota-Baxter
operators whereas in the literature they are generally called
Baxter operators. This serves mainly to distinguish them clearly
from the Yang-Baxter family of objects -by the way both Baxter are
different.} of weight \mkb{$\lambda \in \K$} fulfil the so called
Rota-Baxter relation which may be regarded as one possible
generalization of the standard shuffle relation \cite{Rota1, LG}.
They appeared for the first time in the work of the mathematician
G.~Baxter \cite{B} and were then intensively studied by
\mkb{F. V. Atkinson} \cite{A}, J.~B.~Miller \cite{M}, G.-C.~Rota
\cite{Rota2,Rota3}, P.~Cartier \cite{Car} and others, while more
recently they reappeared in the work of L.~Guo \cite{LG}. \\
There is a lot more to say about Rota-Baxter operators and the
Rota-Baxter relation which will be part of a series of future work
devoted to certain other aspects of Rota-Baxter operators and the
Rota-Baxter relation. Especially the class of idempotent
Rota-Baxter operators showed up to be of importance recently with
regards to the Hopf algebraic background of renormalization, in
particular
in the context of the minimal subtraction scheme and the Riemann-Hilbert
problem \cite{CK1,CK2}.\\
By Loday-type algebras we mean  dendriform di- and trialgebras
\cite{JLL, JLLR}. Such algebras are equipped with two respectively
three algebra compositions fulfilling  certain relations.\\
In an interesting and inspiring work M.~Aguiar \cite{MA} showed,
beside other results, that the class of Rota-Baxter operators of
weight $\lambda=0$ defined on an associative $\K$-algebra $\A$
allows one to define a dendriform dialgebra due to the Rota-Baxter
relation which reduces in the case of zero weight ($\lambda=0$) to a
mere
shuffle relation.\\
Here we would like to show that one can extend the aforementioned
result of Aguiar to an associative $\K$-algebra equipped with a
general Rota-Baxter operator of weight $\lambda \in \K$ fulfilling
the Rota-Baxter relation. Let us call such an associative
$\K$-algebra containing a Rota-Baxter operator a Rota-Baxter
algebra. We will see that the most natural dendriform structure on
a Rota-Baxter algebra is a dendriform trialgebra one.
The connection between Rota-Baxter algebras and Loday-type
algebras provides a rich class of examples for the latter. Our
observation was inspired by a paper of M. A. Semenov-Tian-Shansky
on the classical $r$-matrix and the so called modified classical
Yang-Baxter equation defined therein \cite{STS}.\\
This paper is organized as follows. In the next section we first
give a brief sketch of the concept of Rota-Baxter algebras mainly
relying on the articles by L.~Guo \cite{LG} and G.-C.~Rota
\cite{Rota1}. Section three contains the definitions of dendriform
di- and trialgebras as they can be found in the exhaustive work of
J.-L. Loday and collaborators \cite{JLL, JLLR}. In section four we
extend the observation of M. Aguiar \cite{MA} to general
Rota-Baxter operators including the link to dendriform
trialgebras. This provides a whole new class of examples for these
algebraic structures. The last section ends with a short summary
and an outlook to the forthcoming work in progress.
%
%
%
%
\section{The Rota-Baxter Relation}
Let $\A$ be an associative $\K$-algebra. $\K$ is supposed to be a field
$(\C \makebox{ or } \R)$.
The linear operator $R:\A  \to \A$ must fulfil the following relation:
\beq
      R(x)R(y) + \lambda R(x\: y) = R(R(x)\: y + x\: R(y)),\;\; x,y \in \A. \label{RBR}
\eeq
The constant $\lambda \in \K$ is fixed once and for all and is
called the weight. This set-up can easily be generalized as,
for instance, in \cite{LG}. We call the relation (\ref{RBR}) the
Rota-Baxter relation (RBR) and the operator $R$ is called
Rota-Baxter operator (RBO) of weight $\lambda \in \K$.
Let us call the tuple $\A_R:=(\A;\; R)$ a Rota-Baxter $\K$-algebra of
weight $\lambda \in \K$.\\
The case $\lambda = 0$:
\beq
      R(x)R(y) = R(R(x)\: y + x\: R(y)),\;\; x,y \in \A \label{tRBR}
\eeq
may easily be identified as a shuffle relation as one can see it for instance by defining the RBO $R$ as the
integral operator on a well chosen function algebra, it then reflects the rule of integration by parts:
\beq
      R[f](x):=\int_{0}^{x}f(y)\:dy.
\eeq
For the case of an arbitrary $\lambda \in \K$ the relation (\ref{RBR}) should therefore be regarded as a possible
generalization of the shuffle relation. Of great interest is the class of idempotent RBOs $(R=R^2)$ on which we will
comment in the last section.\\
For the rest of this paper we concentrate on the natural case of $\lambda = + 1$ which can always be
achieved by a normalization of \mkb{$R \to \lambda^{-1}R,\;\lambda \ne 0$}.
Nevertheless we will use the phrase {\emph{Rota-Baxter operator of arbitrary weight $\lambda$}},
but ignore the $\lambda$s in most of the equations.\\
As Atkinson showed in \cite{A}, a linear operator $R$ from $\A$ to $\A$ satisfying the RBR
is equivalent to the fact that $\A$ is a subdirect difference in the sense of Birkhoff of
two subalgebras \mkb{$\P, \F \subset \A$}.\\
Suppose $R$ fulfils the RBR of weight $\lambda =1$:
\beq
      R(x)R(y) + R(x\: y) = R(R(x)\: y + x\: R(y)), \label{1RBR}
\eeq
the same is then also true for the "opposite" operator \mkb{$R^{-}:=1 - R$}.\\
We would like to present two examples of Rota-Baxter algebras. The
first one was introduced by Miller in \cite{M}. Let $\A$ be a
finitedimensional $\K$-vector space with basis $e_1,\dots,e_n$ and
make it into an associative algebra by defining the product
componentwise on the column matrices of \mkb{$n=s+t$} components:
\beq
a,b \in \A,\; a_i, b_j \in \K,\; \;\; a \cdot b = \sum_{i=1}^{n} a_i e_i \cdot \sum_{j=1}^{n} b_j e_j :=  \sum_{i=1}^{n}(a_i b_i)e_i,\;\;
\eeq
Then the following matrix $R$ defines a RBO of weight $1$:
\beq
 R:=\left( \matrix{
                    S_s & 0 \cr
                    0 & T_t \cr
                             } \right),
\eeq
\beq
  S_s:=\left( \matrix{
                    1 & 1  & \ldots & 1      \cr
                    0 & 1  & \ldots & 1      \cr
                    \vdots & \ddots & \ddots & \vdots\cr
                    0 & \ldots & 0&  1 \cr
                             } \right)_{s \times s},
                             \;\;\;
   T_t:=\left( \matrix{
                     0 &  0  & \ldots & 0    \cr
                    -1 &  0  & \ldots & 0    \cr
                \vdots & \ddots & \ddots & \vdots \cr
                    -1 & \ldots & -1 & 0  \cr
                             } \right)_{t \times t}
\eeq
Another example is provided by the algebra of Laurent polynomials.
The (idempotent) RBO is now given by the projector $R_{ms}$:
\beq
 R_{ms}(\sum_{i=-m}^{\infty}c_i z^i) := \sum_{i=-m}^{-1}c_i z^i. \label{Rms}
\eeq
$$
x:=\sum_{i=-m}^{\infty}a_i z^i, \;\; y:=\sum_{j=-n}^{\infty} b_j z^j
$$
\beqn
 R_{ms}\big( R_{ms}(x)y+xR_{ms}(y)-xy \big)&=& R_{ms} \big(\sum_{i=-m}^{-1}a_i z^i \: \sum_{j=-n}^{\infty} b_j z^j +         \nonumber\\
                                           & &        \sum_{i=-m}^{\infty}a_i z^i \: \sum_{j=-n}^{-1} b_j z^j  -
                                                      \sum_{i=-m}^{\infty}a_i z^i \:\sum_{j=-n}^{\infty} b_j z^j \big)       \nonumber\\
\eeqn
\beqn
                                           &=& R_{ms} \big(\sum_{i=-m}^{\infty}a_i z^i \: \sum_{j=-n}^{-1} b_j z^j -
                                                           \sum_{i=0}^{\infty}a_i z^i \:\sum_{j=-n}^{\infty} b_j z^j \big)   \nonumber\\
                                           &=& R_{ms} \big(\sum_{i=-m}^{-1}a_i z^i \: \sum_{j=-n}^{-1} b_j z^j -
                                                           \sum_{i=0}^{\infty}a_i z^i \:\sum_{j=0}^{\infty} b_j z^j \big)    \nonumber\\
                                           &=& R_{ms} \big(\sum_{i=-m}^{-1}a_i z^i \: \sum_{j=-n}^{-1} b_j z^j \big) -
                                               R_{ms} \big(\sum_{i=0}^{\infty}a_i z^i \:\sum_{j=0}^{\infty} b_j z^j \big)    \nonumber\\
                                           &\stackrel{(\ref{Rms})}{=}&
                                                        R_{ms} \big(\sum_{i=-m}^{-1}a_i z^i \: \sum_{j=-n}^{-1} b_j z^j \big)\nonumber\\
                                           &=&\sum_{i=-m}^{-1}a_i z^i \;  \sum_{j=-n}^{-1} b_j z^j                           \nonumber\\
                                           &=&  R_{ms}(x)R_{ms}(y)            \label{cal}
\eeqn
With $R_{ms}$ the opposite operator $R^{-}_{ms}:=1-R_{ms}$ is also a RBO:
\beq
 R^{-}_{ms}(\sum_{i=-m}^{\infty}c_i z^i) := \sum_{i=0}^{\infty}c_i z^i.
\eeq
The projector $R_{ms}$ is used in some renormalization procedure of QFT which is called the minimal
subtraction scheme and where $R_{ms}$ is a so called renormalization map.
It is intimately related to the Riemann-Hilbert problem as was shown in \cite{CK1, CK2}.\\
Replacing the $-1$ on the rhs in (\ref{Rms}) by $0$ also gives a RBO.
For a general \mkb{$r \in \Z \backslash\{-1,0\}$}, noted by $R_{r}$:
\beq
 R_{r}(\sum_{i=-m}^{\infty}c_i z^i) := \sum_{i=-m}^{r}c_i z^i
\eeq
the RBR (\ref{1RBR}) does not hold, as one can see by the
following argument. For $r > 0$ one can show that one gets on the
rhs of (\ref{cal}) a polynomial of order $2r$ whereas on the lhs one
only gets a polynomial of order $r$.
For $ r<-1 $ the same argument applies.\\
A more detailed presentation of Rota-Baxter algebras will be given elsewhere.\\
Inspired by the work of Semenov-Tian-Shansky \cite{STS} we define now the following
new operator on the Rota-Baxter algebra $\A_R$, $R$ of weight $\lambda$:
\beq
  B_{\lambda}:=\lambda - 2R \label{B-OP}
\eeq
which we will call modified Rota-Baxter operator of weight $\lambda$ and which fulfils the relation:
\beq
      B_{\lambda}(x)B_{\lambda}(y)  = B_{\lambda}(B_{\lambda}(x)\: y + x\: B_{\lambda}(y)) - \lambda^2 x\:y. \label{gmaYBR}
\eeq
\beqn
 B_{\lambda}(x)B_{\lambda}(y) &=& \lambda^2 xy -2\lambda(R(x)y + xR(y)) +4R(x)R(y)                                       \nonumber\\
          &\stackrel{(\ref{1RBR})}{=}& 2\lambda^2xy -\lambda^2 xy - 2\lambda (R(x)y + xR(y)) +                           \nonumber\\
          & & \phantom{a}\hspace{4cm}4 (R(R(x)\: y + x\: R(y)) - \lambda R(x\: y))                                       \nonumber\\
          &=& (\lambda-2R)(\:2\lambda xy -2  R(x)y - 2 xR(y)\:) - \lambda^2 xy                                           \nonumber\\
          &=& (\lambda -2R)(\:(\lambda x-2R(x))y + x(\lambda y-2R(y))\:) - \lambda^2 xy                                  \nonumber\\
          &=& B_{\lambda}(B_{\lambda}(x)y + xB_{\lambda}(y)) - \lambda^2 xy                                              \nonumber\\
          & &\phantom{a}\hspace{9cm}\blacksquare                                                                         \nonumber
\eeqn
Equation (\ref{gmaYBR}) is called modified Rota-Baxter relation.
As we already said before when normalizing the RBO of weight $\lambda \ne 0$
to $\lambda^{-1}R$ it holds the RBR (\ref{1RBR}).
The operator \mkb{$B:=1-2R$} then fulfils the relation:
\beq
      B(x)B(y)  = B(B(x)\: y + x\: B(y)) - x\:y. \label{maYBR}
\eeq
Whereas the opposite operator \mkb{$\tilde{B}:=1+2R$} fulfils (\ref{maYBR})
if the RBO $R$ is of weight \mkb{$\lambda=-1$}.\\
In the Lie-algebraic context of \cite{STS} this relation is called (the operator form of the)
modified classical Yang-Baxter equation. Let us remark here that in a Lie-algebraic context
the relation (\ref{1RBR}) is called (operator form of the) classical Yang-Baxter equation \cite{BD} .
Since we work here in the realm of associative algebras we will call expression (\ref{maYBR})
the modified associative classical Yang-Baxter relation (maCYBR).
We follow hereby the terminology introduced by Aguiar in \cite{MA, MA2} who defined on an associative algebra $\A$
an associative analog of the classical Yang-Baxter equation for the $r$-matrix \mkb{$r \in \A \otimes \A$}:
\beq
     aCYBE(r):= r_{13}r_{12} - r_{12}r_{23} + r_{23}r_{13}=0 \label{aCYBE}
\eeq
We would like to underline here that the RBR (\ref{1RBR}) with
respect to the equation (\ref{aCYBE}) may be interpreted in the
same way as it is done in the Lie-algebraic context of the CYBE in
\cite{BD}. This interesting link of the (associative and modified
associative) classical Yang-Baxter relation to the realm of
Rota-Baxter operators is part of work in progress \cite{KEF}.
%
%
%
%
%
\section{The Dendriform Di- and Trialgebra}
We will give here the definitions of a dendriform di- and trialgebra following
the work of Loday and Loday and Ronco \cite{JLL, JLLR}. Let $\A$ be a $\K$-vector
space equipped with the following two binary compositions:
\beqn
\prec : \A \otimes \A \to \A \nonumber\\
\succ : \A \otimes \A \to \A \nonumber
\eeqn
which are supposed to hold the so called dendriform dialgebra relations \cite{JLL}:
\beqn
(a \prec b ) \prec c  &=&  a \prec (b \prec c) + a \prec (b  \succ c)    \nonumber\\
 a \succ (b  \prec c) &=& (a \succ b) \prec c                            \nonumber\\
a \succ ( b  \succ c)  &=& (a \prec b) \succ c + (a \succ b)  \succ c    \nonumber\\
                                                                         \label{Ddi}
\eeqn
The triple \mkb{$(\A, \prec, \succ)$} is then called a dendriform dialgebra.\\
We come now to the trialgebra structure.
As before let $\A$ be a $\K$-vector space equipped with the three binary compositions:
\beqn
\prec : \A \otimes \A \to \A    \nonumber\\
\succ : \A \otimes \A \to \A    \nonumber\\
\cdot  : \A \otimes \A \to \A   \nonumber \eeqn which are supposed
to satisfy the following  relations, the so called dendriform
trialgebra axioms \cite{JLLR}:
\beqn
(a \prec b) \prec c   &=&  a \prec (b \prec c + b \succ c + b\cdot c)                \label{tri1}\\
(a \succ b)  \prec c  &=&  a \succ (b \prec c)                                       \label{tri2}\\
 a \succ (b \succ c)  &=& (a \prec b + a \succ b + a\cdot b ) \succ c                \label{tri3}\\
(a \prec b) \cdot c  &=&  a \cdot (b \succ c)                                        \label{tri4}\\
(a \succ b) \cdot c  &=&  a \succ   (b \cdot c)                                      \label{tri5}\\
(a \cdot b) \prec c  &=&  a \cdot (b \prec c)                                        \label{tri6}\\
(a \cdot b) \cdot c  &=&  a \cdot (b \cdot c)                                        \label{tri7}
\eeqn
We also define a fourth multiplication on $\A$:
\beq
*:\A \otimes \A \to \A,\;\; a*b := a \prec b + a \succ b + a\cdot b  \label{tri8}
\eeq
which shows to be an associative binary composition on $\A$:
\beqn
(a * b ) * c  &=& (a \prec b + a \succ b + a\cdot b)*c                                                                 \nonumber\\
              &=& (a \prec b)\prec c +
                                 (a \prec b) \succ c +
                                               (a \prec b)\cdot c +                                                    \nonumber\\
              & &  \;\;\;\;\;\;  (a \succ b) \prec c + (a \succ b)\succ c + (a \succ b)\cdot c +                       \nonumber\\
              & &  \;\;\;\;\;\;\;\;\;\;\;\;\;\;\;\;\;\;  (a \cdot b) \prec c + (a \cdot b)\succ c + (a \cdot b)\cdot c \nonumber\\
              &\stackrel{(\ref{tri1},\ref{tri3}, \ref{tri4})}{=}& a \prec (b * c) +
                                                                              a \succ (b \succ c) +
                                                                                                  a\cdot (b \succ c) + \nonumber\\
              & &  \;\;\;\;\;\;  (a \succ b) \prec c + (a \succ b)\cdot c +                                            \nonumber\\
              & &  \;\;\;\;\;\;\;\;\;\;\;\;\;\;\;\;\;\;\;\;\;  (a \cdot b) \prec c + (a \cdot b)\cdot c                \nonumber\\
              &\stackrel{(\ref{tri2},\ref{tri5})}{=}& a \prec (b * c) +
                                                                        a \succ (b \prec c + b \succ c + b \cdot c) +  \nonumber\\
              & &  \;\;\;\;\;\;  a\cdot (b \succ c) + (a \cdot b) \prec c + (a \cdot b)\cdot c                         \nonumber\\
              &\stackrel{(\ref{tri6})}{=}& a \prec (b * c) + a \succ (b * c) +                                         \nonumber\\
              & &  \;\;\;\;\;\;  a \cdot (b \succ c + b \prec c) + (a \cdot b)\cdot c                                  \nonumber\\
              &\stackrel{(\ref{tri7})}{=}& a*(b*c)
\eeqn
We will not go into any details with respect to these algebraic
structures which can be found in the before-mentioned literature.
Loday et al. give several examples of dendriform di- and trialgebras in \cite{JLL, JLLR}.\\
In the following section we will show that Rota-Baxter operators of arbitrary weight $\lambda$ respectively
Rota-Baxter algebras provide another class of interesting examples for these two algebraic structures.
%
%
%
%
%
\section{Dendriform Di- and Trialgebra structures on Rota-Baxter algebras}
In \cite{MA} Aguiar observed that on a Rota-Baxter algebra with RBO of weight $\lambda=0$, \mkb{$(\A; \;R)$},
the following two binary compositions:
\beqn
\prec : \A \otimes \A \to \A \nonumber\\
    a \prec b &:=& aR(b)     \nonumber\\
\succ : \A \otimes \A \to \A \nonumber\\
    a \succ b &:=& R(a)b     \nonumber
\eeqn
fulfil the dendriform dialgebra relations (\ref{Ddi}) and therefore
$(\A,\prec,\succ)$ is a $\K$-dendriform dialgebra.\\
This observation may be extended to general Rota-Baxter algebras with
a RBO of arbitrary weight $\lambda \in \K$ by using the modified
Rota-Baxter operator \mkb{$B_{\lambda}=\lambda-2R$}.
Then the two binary compositions:
\beqn
\prec : \A \otimes \A \to \A              \nonumber\\
    a \prec b &:=& aB(b) - \lambda ab     \nonumber\\
\succ : \A \otimes \A \to \A              \nonumber\\
    a \succ b &:=& B(a)b + \lambda ab     \nonumber
\eeqn
make $(\A,\prec,\succ)$ a dendriform dialgebra as we will show now.
We mentioned above that one can always normalize the RBO $R$ to
\mkb{$\hat{R}:=\lambda^{-1}R, \; \lambda \ne 0$} to fulfil the
RBR for the weight $+ 1$ so that the operator \mkb{$B=1-2R$}
holds the relation:
\beq
      B(x)B(y)  = B(B(x)\: y + x\: B(y)) - x\:y. \label{maYBE}
\eeq
Of course we could have defined them also by the two RBOs  $R$ and $R^{-}$, \mkb{$a \prec b = -2aR(b)$}
and \mkb{$a \succ b:=2R^{-}(a) b$}, \mkb{$R^{-}:=1-R$} \footnote{From this point of view it would be more
convenient to use a RBO of weight $\lambda=2$.}.
\beqn
(a \prec b ) \prec c  &=& aB(b)B(c) - abB(c) - aB(b)c + abc                                                             \nonumber\\
                      &=& a(B(b)B(c) + bc)- abB(c) - aB(b)c                                                             \nonumber\\
                      &\stackrel{(\ref{maYBE})}{=}& aB(B(b)c) + aB(bB(c))- abB(c) - aB(b)c                              \nonumber\\
                      &=& aB(B(b)c) + aB(bB(c))- abB(c) - aB(b)c +                                                      \nonumber\\
                      & & \phantom{a} \;\;\;\;\;\;\;\;\;\;\;\;\;\;\;\;\;\;\;\;\;\;\;\;\;\; + abc -abc + aB(bc) - aB(bc) \nonumber\\
                      &=& a \prec (bB(c) -bc) + a \prec (B(b)c+bc)                                                      \nonumber\\
                      &=& a \prec (b \prec c) + a \prec (b  \succ c)                                                    \nonumber\\[0.3cm]
a \succ (b \prec c)  &=& B(a)bB(c) - B(a)bc + abB(c) -abc            \nonumber\\
                     &=& (B(a)b + ab)B(c) - (B(a)b + ab)c            \nonumber\\
                     &=& (a \succ b) \prec c                         \nonumber\\[0.3cm]
a \succ ( b \succ c) &=& B(a)B(b)c + B(a)bc + aB(b)c +abc                                                 \nonumber\\
                     &\stackrel{(\ref{maYBE})}{=}& (B(B(a)b) + B(aB(b) -ab)c  + B(a)bc + aB(b)c +abc      \nonumber\\
                     &=& (aB(b) - ab) \succ c + (B(a)b + ab) \succ c                                      \nonumber\\
                     &=& (a \prec b) \succ c + (a \succ b) \succ c                                        \nonumber\\
                     & &\phantom{a}\hspace{9cm}\blacksquare                                               \nonumber
\eeqn
We mention as an aside that for an idempotent RBO the RBR
(\ref{1RBR}) is fulfilled on both compositions $(\prec,\succ)$:
\beqn
      R(x)\prec R(y) + R(x \prec y) &=& R(R(x) \prec y + x \prec R(y)) \nonumber\\
      R(x)\succ R(y) + R(x \succ y) &=& R(R(x) \succ y + x \succ R(y)) \nonumber
\eeqn
Regarding the result of Aguiar it seems to us that the most natural dendriform structure
on a Rota-Baxter algebra of weight \mkb{$\lambda \ne 0$} is the trialgebra one.
On the Rota-Baxter algebra $(\A;\:R)$ we denote the multiplication now by
\mkb{$a \cdot b \in \A, \;\; a,b \in \A$}.
The RBO is supposed to be of weight $\lambda = -1$ just for reasons of clarity
with regard to the trialgebra axioms in (\ref{tri1}-\ref{tri7}).\\
We define the following two binary compositions:
\beqn
\prec : \A \otimes \A \to \A      \nonumber\\
    a \prec b &:=& a \cdot R(a)   \nonumber\\
\succ : \A \otimes \A \to \A      \nonumber\\
    a \succ b &:=& R(a) \cdot b   \nonumber
\eeqn
The fourth composition looks as follows:
\beq
*:\A \otimes \A \to \A,\;\; a*b := a \cdot R(b) + R(a) \cdot b + a \cdot b.  \label{drp}
\eeq
We show now that these multiplications hold the axioms in (\ref{tri1}-\ref{tri7}) and that (\ref{drp}) is associative.
This makes \mkb{$(\A,\prec,\succ,\cdot)$} a dendriform trialgebra.\\
Relations (\ref{tri2}) and (\ref{tri4}-\ref{tri7}) are easy to show by the associativity of the Rota-Baxter algebra.
The axioms (\ref{tri1},\ref{tri3}) follow from the RBR (\ref{1RBR}):
\beqn
(a \prec b ) \prec c  &=& a\cdot R(b)\cdot R(c)                                                          \nonumber\\
                      &\stackrel{(\ref{1RBR})}{=}& a \cdot R( R(b)\cdot\: c + b\cdot R(c) + b \cdot c )  \nonumber\\
                      &=& a\cdot R( b \succ c + b \prec c + b \cdot c )                                  \nonumber\\
                      &=& a \prec ( b \succ c + b \prec c + b \cdot c )                                  \nonumber\\[0.3cm]
%
%
a \succ ( b \succ c) &=& R(a)\cdot R(b) \cdot c                                                       \nonumber\\
                     &\stackrel{(\ref{1RBR})}{=}& R( R(a)\cdot b + a\cdot R(b) + a \cdot b )\cdot c   \nonumber\\
                     &=& R( a \succ b + a \prec b + a \cdot b) \cdot c                                \nonumber\\
                     &=&  ( a \succ b + a \prec b + a \cdot b) \succ c                                \nonumber\\
\eeqn
We are left to show the associativity of the fourth composition $*:\A \otimes \A \to \A$:
\beqn
(a * b ) * c  &=& (R(a)\cdot b + a \cdot R(b) + a\cdot b)*c                                                          \nonumber\\
              &\stackrel{(\ref{1RBR})}{=}& R(a)\cdot R(b)\cdot c +                                                   \nonumber\\
              & & \phantom{a} \hspace{1cm} R(a)\cdot b \cdot  R(c) + a \cdot R(b)\cdot R(c) + a\cdot b\cdot R(c) +   \nonumber\\
              & & \phantom{a} \hspace{3cm} + R(a)\cdot b \cdot  c + a \cdot R(b)\cdot c + a\cdot b\cdot c            \nonumber\\
              &\stackrel{(\ref{1RBR})}{=}& R(a) \cdot (R(b)\cdot c + b \cdot  R(c) + b \cdot  c)                     \nonumber\\
              & & \phantom{a} \hspace{1cm} + a \cdot R(R(b)\cdot c + b \cdot  R(c) + b\cdot c)                       \nonumber\\
              & & \phantom{a} \hspace{3cm} + a \cdot (R(b)\cdot c + b \cdot  R(c) + b\cdot c)                        \nonumber\\                                         &=& a * (R(b)\cdot c + b \cdot  R(c) + b\cdot c)                                                      \nonumber\\
              &=& a * ( c * b )                                                                                      \nonumber\\
                     & &\phantom{a}\hspace{9cm}\blacksquare                                                          \nonumber
\eeqn
Rota-Baxter algebras may thus be equipped with dendriform di- and
trialgebra structures. It is interesting to see this dendriform
structures on Rota-Baxter algebras with respect to Atkinson's
theorem mentioned above.\\[0.1cm]
Finally we would like to comment briefly on the homogeneous version of the
Rota-Baxter relation of weight $\lambda$ defined on an associative $\K$-algebra $\A$:
\beq
      R(x)R(y) + \lambda R^2(x\: y) = R(R(x)\: y + x\: R(y)),\;\; x,y \in \A, \label{aNR}
\eeq
which is for $\lambda=+1$ an associative analog of the Nijenhuis relation. Let us
call an operator fulfilling (\ref{aNR}) a Nijenhuis operator of weight $\lambda$.
Now the arbitrariness with respect to the parameter $\lambda$ is lost.
Let us remark here, that having an idempotent RBO $R$ of weight $+1$, it is possible to define arbitrarily
many Nijenhuis operators \mkb{$N_{\alpha}:= R - \alpha R^{-}$}, \mkb{$\alpha \in \K $}
of weight $\lambda=+1$.\\
Exactly for a Nijenhuis operator $N$ fulfilling ({\ref{aNR}) for $\lambda=+1$ the product:
\beq
*:\A \otimes \A \to \A,\;\; a*b := a \cdot N(b) + N(a) \cdot b - N(a \cdot b).  \label{np}
\eeq
is again associative and $a \prec b:= aN(b),\;\; a \succ b:=N(a)b$ hold the dendriform
trialgebra axioms.
%
%
%
%
%
\section{Summary and Outlook}
We presented briefly the concept of Rota-Baxter algebras and
introduced some new operators which we called modified Rota-Baxter
operator. It was shown that these operators fulfil a new equation
which we called the modified associative classical Yang-Baxter relation.
We mentioned also the intimate relation of the Rota-Baxter relation
and the associative analog of the classical Yang-Baxter equation.\\
After giving the definitions of dendriform di- and trialgebras we used the modified Rota-Baxter
operators to extend the result of Aguiar to associative $\K$-algebras equipped with a Rota-Baxter
operator of arbitrary weight $\lambda$. The trialgebra structure shows to be the most natural
dendriform structure on a Rota-Baxter algebra of arbitrary weight.\\
The structure implied on an associative $\K$-algebra by the
general Rota-Baxter relation is quite astonishing and requires
further investigations. We remark here briefly that on a
Rota-Baxter algebra equipped with the associated Lie-bracket the
modified classical Yang-Baxter relation is fulfilled and one can define a
pre-Lie structure related to the Jordan product and the modified
Rota-Baxter operator.\\
The Rota-Baxter relation is also of importance with respect to
multiple-zeta-values (MZVs) and a possible q-deformation of them.
All these structures relate to quantum field theory \cite{CK1, CK2, DK1}
and we expect a further study of Rota-Baxter operators to be of significance
for the understanding of such theories.\\[0.5cm]
{\emph{Acknowledgments}}:\\
A special vote of thanks goes to Prof. D. Kreimer for helpful comments concerning this work.
The author is supported by the Evangelisches Studienwerk Villigst.
%
%

%
%

%
%
%
\end{document}